\documentclass[fleqn,usenatbib,letters]{mnras}

\usepackage{graphicx}
\usepackage{amssymb}
\usepackage{epstopdf}
\usepackage{amsmath}
\usepackage{amssymb}
\usepackage{comment,xcolor}
\usepackage{hyperref}
\usepackage{gensymb}
\usepackage{xspace}
\usepackage{array}
\usepackage{cuted}
\usepackage{orchidlink}
\usepackage{tikz}
\usepackage{booktabs}  
\usepackage{animate}
\usepackage{media9}
\usepackage{here}
\usepackage{lipsum}
\usepackage{float}
\usepackage{subcaption}  
\usepackage{soul}
\usepackage{here}
\usepackage[colorinlistoftodos,prependcaption,textsize=small]{todonotes}
\usepackage[normalem]{ulem}

\newcommand{\tikzcircle}[2][red,fill=red]{\tikz[baseline=-0.5ex]\draw[#1,radius=#2] (0,0) circle ;}%

\definecolor{dg}{rgb}{0.0, 0.6, 0.1}
\newcommand{\Andrew}[1]{\textcolor{dg}{#1}}

\newcommand{\Vasu}[1]{{\color{black}#1}}

\DeclareRobustCommand{\VAN}[3]{#2}
\let\VANthebibliography\thebibliography
\def\thebibliography{\DeclareRobustCommand{\VAN}[3]{##3}\VANthebibliography}

\title{Influence of the Galactic Halo on the UHECR Multipoles}

\author[V.~Shaw et al.]{
Vasundhara~Shaw$^{1,2}$\thanks{E-mail: vasundhara.shaw@manchester.ac.uk}\orcidlink{0000-0002-5824-7191},
Arjen~van~Vliet$^{3}$\orcidlink{0000-0003-2827-3361},
Andrew M. Taylor$^{2}$\orcidlink{0000-0001-9473-4758}
\\
$^{1}$Jodrell Bank Centre for Astrophysics, Department of Physics and Astronomy, The University of Manchester, \\ Manchester, M13 9PL, U.K. \\
$^{2}$Deutsches Elektronen-Synchrotron, Platanenallee 6, 15738 Zeuthen, Germany \\ 
$^{3}$Department of Physics, Khalifa University of Science and Technology, P.O. Box 127788, Abu Dhabi, United Arab Emirates
}
\date{Accepted XXX. Received YYY; in original form ZZZ}

\pubyear{2024}

\begin{document}
\include{introduction}

\maketitle

\begin{abstract}
We examine the effects of a giant magnetized halo around the Galaxy on the angular distribution of the arriving ultra-high energy cosmic rays (UHECR) observed at Earth. We investigate three injection scenarios for UHECRs, and track them through isotropic turbulent magnetic fields of varying strengths in the Galactic halo. We calculate the resultant dipole and quadrupole amplitudes for the arriving UHECRs detected by an observer in the Galactic plane region. We find that, regardless of the injection scenario considered, when the scattering length of the particles is comparable to the size of the halo, the UHECRs skymap resembles a dipole. However, as the scattering length is increased, the dipolar moment always increases, and the quadrupolar moment increases rapidly for two of the three cases considered. Additionally, the quadrupole amplitude is highlighted to be a key discriminator in discerning the origin of the observed dipole. We conclude that, to understand the origin of the UHECR dipole, one has to measure the strength of the quadrupole amplitude as well.
\end{abstract}


\section{Introduction}

The arrival directions of ultra-high energy cosmic rays (UHECRs) at Earth is not entirely isotropic. In 2017, the Pierre Auger Collaboration (PAO) reported the detection of dipolar anisotropy, for UHECRs with energies $\geq 8$~EeV~\citep{Auger_2017}. This result was updated in Refs.~\citet{Auger_2018} and~\citet{Auger_2024} and is now at a significance of $6.8\sigma$ with a dipole amplitude of $\delta = \rm {0.074}^{+0.010}_{-0.008}$ being obtained. The direction of this dipole lies ${\sim}~115\degree$ away from the Galactic centre, suggesting an extragalactic UHECR origin. It is worth noting that this cosmic ray dipole does not align with the CMB dipole or any potential known source.



The Telescope Array collaboration (TA) also searched for a large-scale dipole anisotropy in the northern hemisphere sky~\citep{Abbasi_2020}. They reported a dipole amplitude of $3.3 \pm 1.9\%$ which, at their current level of statistics, is {compatible in amplitude and direction with the dipole} reported by PAO, but also with an isotropic distribution. When the datasets of PAO and TA are combined, a dipole with a significance of ${\sim}~4.2\sigma$ is found~\citep{PierreAuger:2023mvf}, {consistent in amplitude and direction with the dipole} reported by PAO alone. Besides the dipole anisotropies, several indications have been found of intermediate-scale anisotropies at higher energies as well (see e.g.~Refs.~\citet{TelescopeArray:2018rtg, Auger_2022}). However, our focus in this paper is predominantly on the large-scale anisotropy.

The large-scale anisotropy in the arrival directions of UHECRs suggests an anisotropy in the distribution of UHECR sources.
This can be attributed to an inhomogeneous distribution of the local large-scale structure (see e.g.~Refs.~\citet{Lang_2021,Ding:2021emg, Allard:2021ioh, Bister:2023icg}) or the existence of one or more bright local sources (see e.g.~Refs.~\citet{Mollerach:2019wne, Eichmann:2022ias, Mollerach:2024sjd}), combined with UHECR propagation through Galactic (GMFs) and/or extragalactic magnetic fields (EGMFs).



The observed UHECR dipole strength at Earth for cosmic rays with energies above 8~EeV is at the 7\% level. However, this dipole strength is not necessarily the same as the arriving extragalactic dipole amplitude, out at large distances where the influence of the Milky Way's magnetic fields has yet to take effect. The Galactic magnetic fields can affect an external extragalactic dipole ($\delta_{\rm{EG}}$) in two main ways: the large-scale structured component changes the extragalactic direction of $\delta_{\rm{EG}}$ and the small-scale turbulent magnetic fields (with RMS value of $B_{\rm{tur}}$) suppress the strength of $\delta_{\rm{EG}}$~\citep{Zirakashvili_2005}. Furthermore, these effects suggest that a beam (current) of particles from a source, upon propagating through the Galactic magnetic field structure, is also expected to be changed, potentially with the arrival directions being spread out by the turbulent field component, {resulting in the injected power in higher multipoles migrating to lower multipoles.} 

The ability of the Milky Way's turbulent magnetic field to alter the cosmic ray distribution locally observed is dependent on the spatial extent and turbulence level of this field component. Over the last few years, a growing body of evidence has accumulated indicating that a considerable level of thermal gas pressure exists out in the halo of the Milky-Way and other local galaxies of a similar mass, out to at least a distance of the virial radius (see \citet{2022MNRAS.511..843M,2022ApJ...928...14B,Hopkins_2022,2024A&A...690A.267Z}). Assuming that the plasma beta of this halo gas is close to unity, a considerable magnetic field strength is also expected out in this region \citep{2020ApJ...893...82F}. Interestingly, such an inference is consistent with recent radio observations of the circumgalactic medium region (at a distance $\sim 100$~kpc) of cosmologically nearby Galaxies. From such studies, the line-of-sight magnetic field in these regions have been found to have {strengths of a few hundred nG}, using Faraday rotation measurements \citep{Heesen_2023}.

In this work we build upon these recent results by considering the effects of {a giant} turbulent Galactic magnetic field halo on the penetration of cosmic rays into the Galaxy. 
In particular, we investigate the effect of the GMF on the arrival of an extragalactic UHECR large-scale dipole distribution, or an extragalactic UHECR small-scale localized distribution. {Our focus in this paper is on the effect that turbulent GMFs have on the observed amplitude of the dipole and quadrupole.}

\section{Methodology - Simulation setup}
\label{sec:sim_setup}



UHECRs gyrating with a Larmor radius larger than the coherence length of the magnetic field, ie. $r_{L}\gg L_{\rm coh}$, in isotropic turbulent magnetic fields, undergo only small angular deflections of size $\Delta_{\Theta}=L_{\rm{coh}}/r_{L} $ in each coherent magnetic field patch, where $\Theta$ is the angle between the particle's initial and final momentum direction.
For a particle propagating a path length $L$, the number of coherence cells traversed, {$N_{\rm{coh}} = L/L_{\rm{coh}}$}. The standard deviation in the deflection angles from this number of small-scale deflections is, {\ $\sigma_{\Theta} = \sqrt{N_{\rm{coh}}}\Delta_{ \Theta}$}.
After a sufficient number of scatterings, the spread in angles becomes large, $\sigma_{\Theta} \sim \pi$ radians. The large angle scattering length can therefore be estimated to be, $L_{\rm{scat}}\approx (\pi/\Delta_{\Theta})^{2}L_{\rm coh} = \pi^2 \frac{r_{L}^2}{L_{\rm{coh}}}$.
The diffusion coefficient $D \sim L_{\rm{scat}}~\rm{c}/3$, where '$\rm{c}$' is the speed of light in a vacuum. Throughout the text will use the terms "scattering length" and "diffusion coefficient" interchangeably. 

 We use CRPropa~3~\citep{CRPropa_2016, CRPropa_2022} to generate isotropic Kolmogorov turbulent magnetic fields {with a power law index of -5/3 to describe the energy distribution of turbulence modes, with  $L_{\rm min}$ = 450~pc and $L_{\rm max}$ = 4000~pc as the minimum and maximum wavelengths, respectively, and with a coherence length ($L_{\rm{coh}}$) of $\approx$ 1~kpc. With the coherence length of the field fixed for each case, the turbulent field effectively has only one free parameter, the magnitude of the turbulent field strength, $B_{\rm tur}$.} 
 
\begin{figure}
    \centering
    \begin{minipage}[c]{0.49\linewidth}
        \begin{subfigure}[t]{\linewidth}
            \centering
            \includegraphics[width=\linewidth]{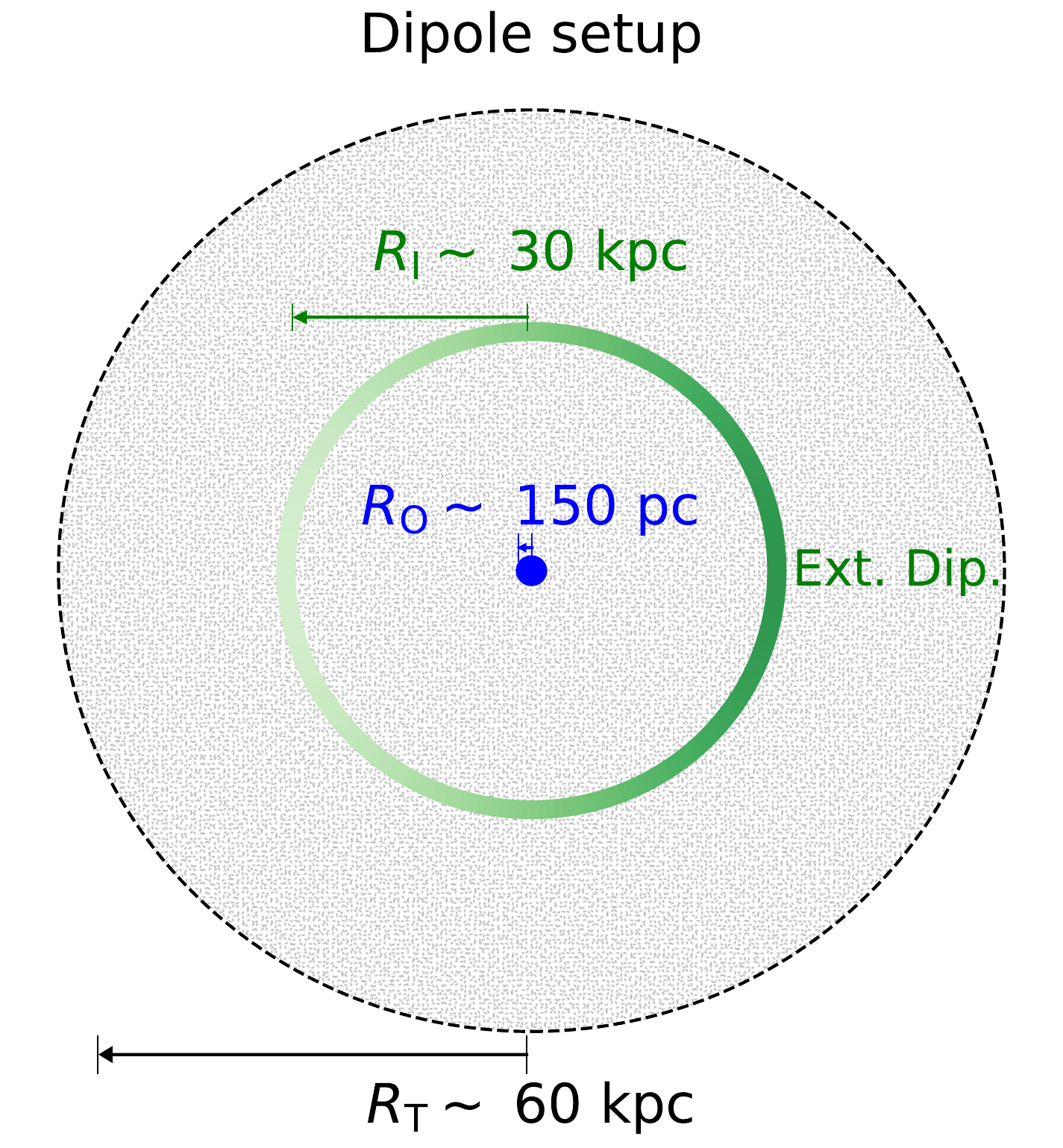}
            \label{fig:cartoons_left}
        \end{subfigure}
    \end{minipage}%
    \begin{minipage}[c]{0.49\linewidth}
        \begin{subfigure}[t]{\linewidth}
            \centering        
            \includegraphics[width=\linewidth]{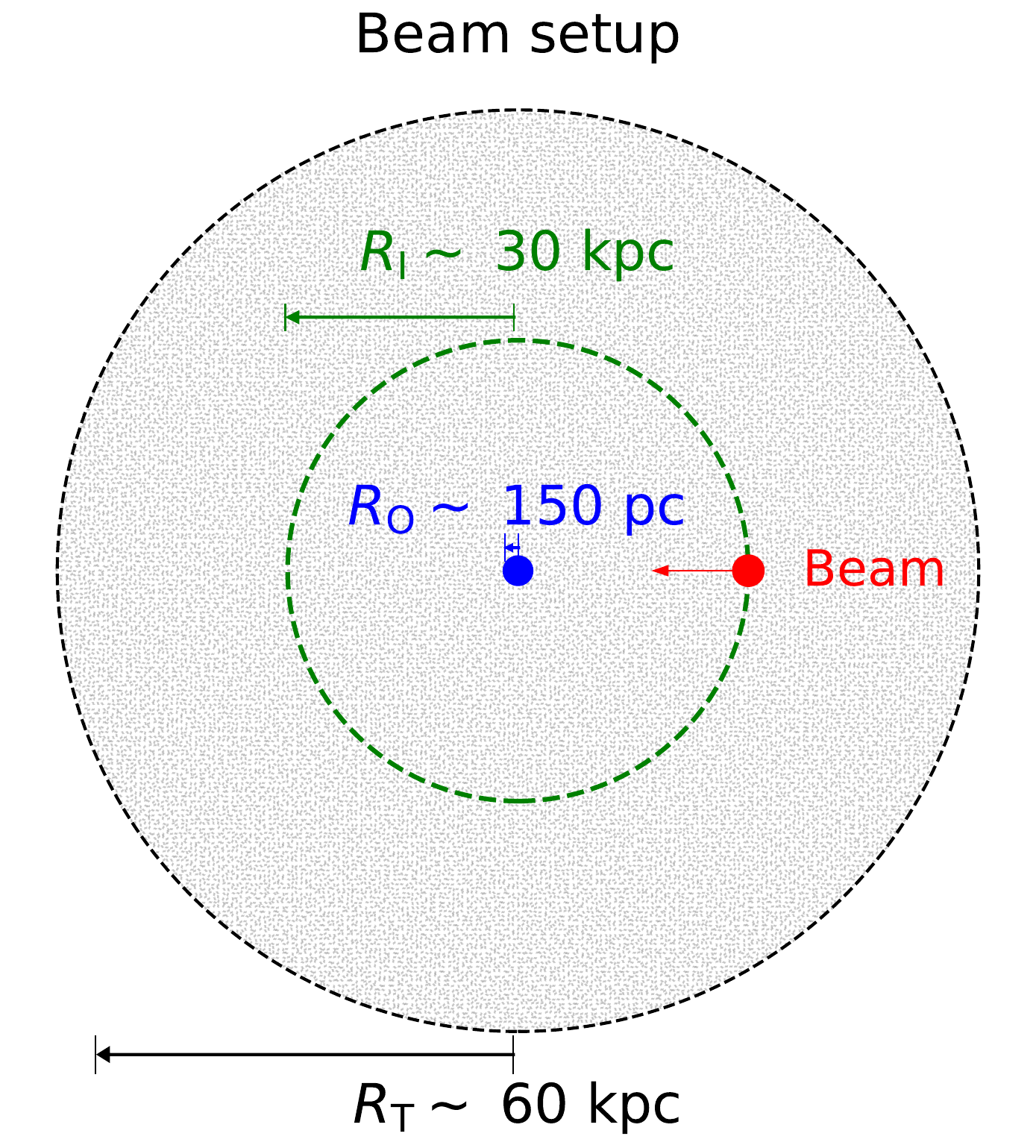}
            \label{fig:cartoons_right}
        \end{subfigure}
    \end{minipage}
    \caption{{Schematic representation of the simulation setup for the dipole and beam setup shown on the \textbf{left} and \textbf{right-hand side}, respectively. In both figures, $R_{\rm T}$, $R_{\rm O}$ and $R_{\rm I}$ are the radii of the termination, observer and injection spheres, respectively. The changing colour of the green ring from right to left indicates the changing strength of the external dipole magnitude, and the single arrow in the beam setup denotes injected UHECRs with the arrow directions pointing inwards in both setups}. {For the extragalactic dipole case, we adopted a Lambert distribution of particle directions.} }
    \label{fig:cartoons}
\end{figure}

 
 The turbulent fields in our setup extend out to the simulation absorption boundary (termination sphere) with radius $R_{\rm T} = 60$~kpc (see Fig.~\ref{fig:cartoons}).
 We note that the length scale values chosen for our setup, specifically $R_{\rm T}$ and $R_{\rm I}$, are chosen for computational rather than physical reasons. The purpose of our work, instead, is a proof of principle of the shielding effect produced by a magnetised halo, rather than an accurate description of the halo. \Vasu{For the giant halo scenario we consider,  cosmic ray transport in the outer Galaxy is no longer 1-D, but 3D, with the cosmic rays achieving a steady state via diffusion into a larger phase space volume region, i.e. no escape boundary is needed for steady state to be achieved in this case. In this regime, the boron to carbon (B/C) ratio is no longer a probe of the true halo size ($R_{\rm T}$), but rather a probe of the size of the region within which the probability of returning back to the solar neighbourhood is of order unity \citep[Sec.~3.1.3]{Hopkins_2022}).}


\subsection{Estimation of the scattering length}
To determine the scattering length ($L_{\rm{scat}}$) of the UHECRs from simulations, we inject these particles in a single direction into the turbulent magnetic field setup described in Sec.~\ref{sec:sim_setup}. 
For each value of $B_{\rm{tur}}$, we ran the simulation until the particles had completely isotropised such that $\langle |{\Theta}| \rangle$, had plateaued at $\approx \pi/2$.
The trajectory length at which $\langle |{\Theta}| \rangle$ plateaus thereby signifies that the particles have lost memory of their injected angular distribution, and is referred to as the particle scattering length $L_{\rm{scat}}$. 
{In order to determine $L_{\rm{scat}}$ for different values of $B_{\rm{tur}}$, we vary $B_{\rm{tur}}$ across 16 logarithmically spaced values from $\approx 0.8~ \text{to}~14$~$\mu$G, whilst maintaining the same coherence length, $L_{\rm{coh}} = 1$~kpc.}


\begin{figure}
    \centering

    \includegraphics[width=1.\linewidth]{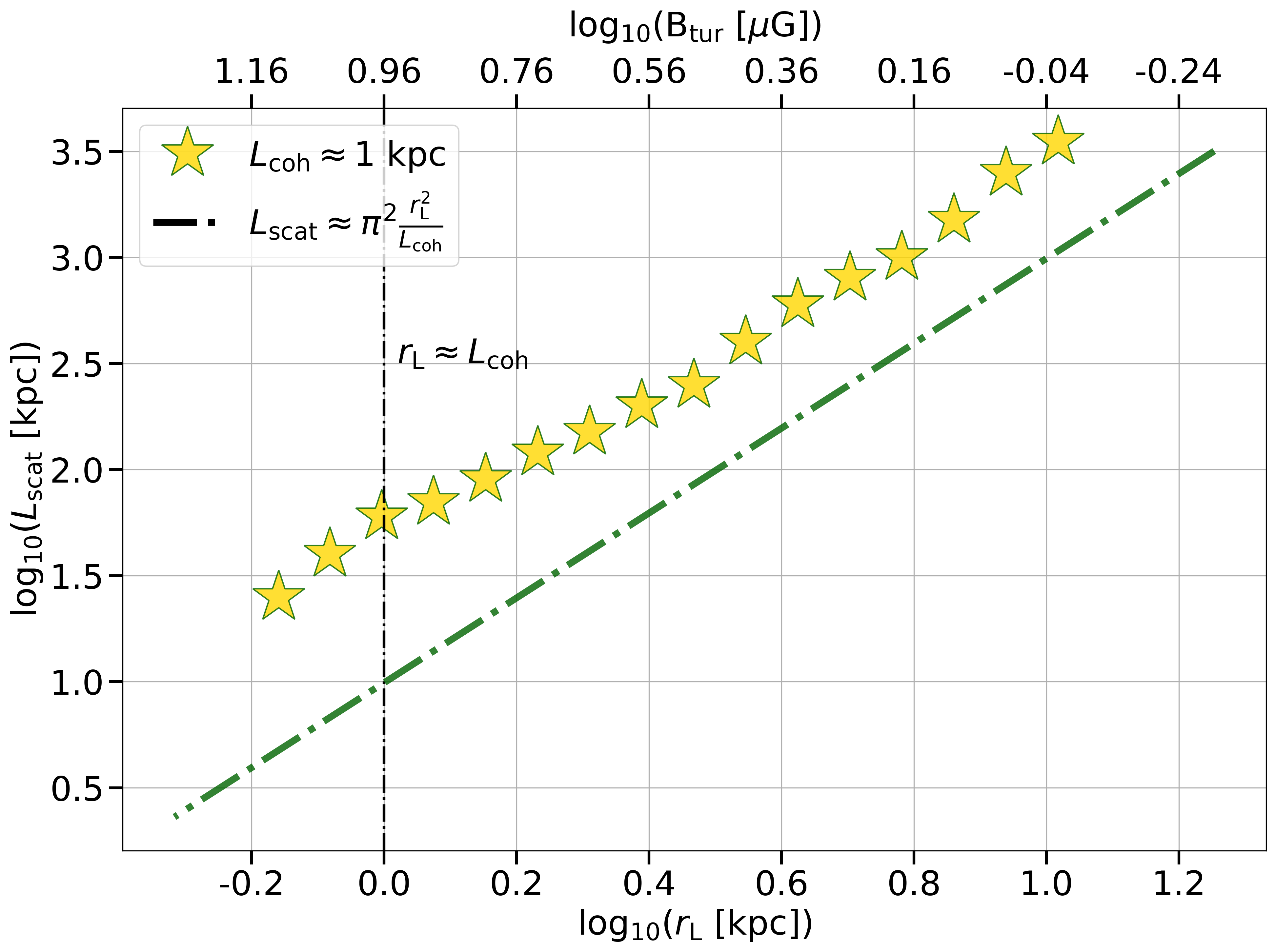}
  
    \caption{{The scattering length ($\star$) plotted as a function of the Larmor radius on the bottom x-axis, and the corresponding turbulent magnetic field strength on the top-secondary x-axis, for 8.5~EeV protons. The green dashed line shows the expected results applicable when $r_{L}> L_{\rm coh}$. Left of the dashed vertical black line shows the region where $r_{\rm{L}} < L_{\rm{coh}}$ and deviation from the small angle scattering regime is expected.}}
    \label{fig:Lscat_Btur}
\end{figure}

Comparing these numerical results with our earlier analytic expectations, for small values of $B_{\rm{tur}}$, such that $r_{L}\gg L_{\rm coh}$, the relation $ L_{\rm{scat}} \approx \pi^2 \frac{r_{L}^2}{L_{\rm{coh}}}$ appears to approximately hold (see Fig.~\ref{fig:Lscat_Btur}). \Vasu{This dependence of the scattering length on the particle Larmor radius, however, is expected to change once the magnetic field strength is further increased, and the UHECR Larmor radius drops below the magnetic field coherence length.

For the case when the Larmor radius is below the coherence length scale, $r_{L}<L_{\rm{coh}}$, resonant interactions with the turbulent modes become the dominant  interaction of the UHECR with the turbulence. The transition to this regime is indicated by the vertical black line in Fig.~\ref{fig:Lscat_Btur}. Due to the limited turbulence dynamical range that we simulate, this regime cannot be strongly probed by our simulations.}


\subsection{Calculation of UHECR dipole amplitudes through turbulent magnetic fields}
\label{dipole_amplitudes}
{In order to create synthetic skymaps for an observer we inject $N_{\rm{inj}} \approx 10^{10}$ UHECRs into the setup depicted in Fig.~\ref{fig:cartoons}. This setup ensures that the observed skymaps contain good statistics ($ N_{\rm{obs}} \approx 4 \times 10^{5}$)}. In reality, UHECRs enter the Galaxy from all directions, and what is observed by PAO \citep{Auger_2017,Auger_2022} is a fraction of the UHECRs that hit the observer (Earth). 



\begin{figure*}
    \centering
    \begin{minipage}[t]{0.45\linewidth}
        \begin{subfigure}[t]{\linewidth}
            \centering
            \includegraphics[width=\linewidth]{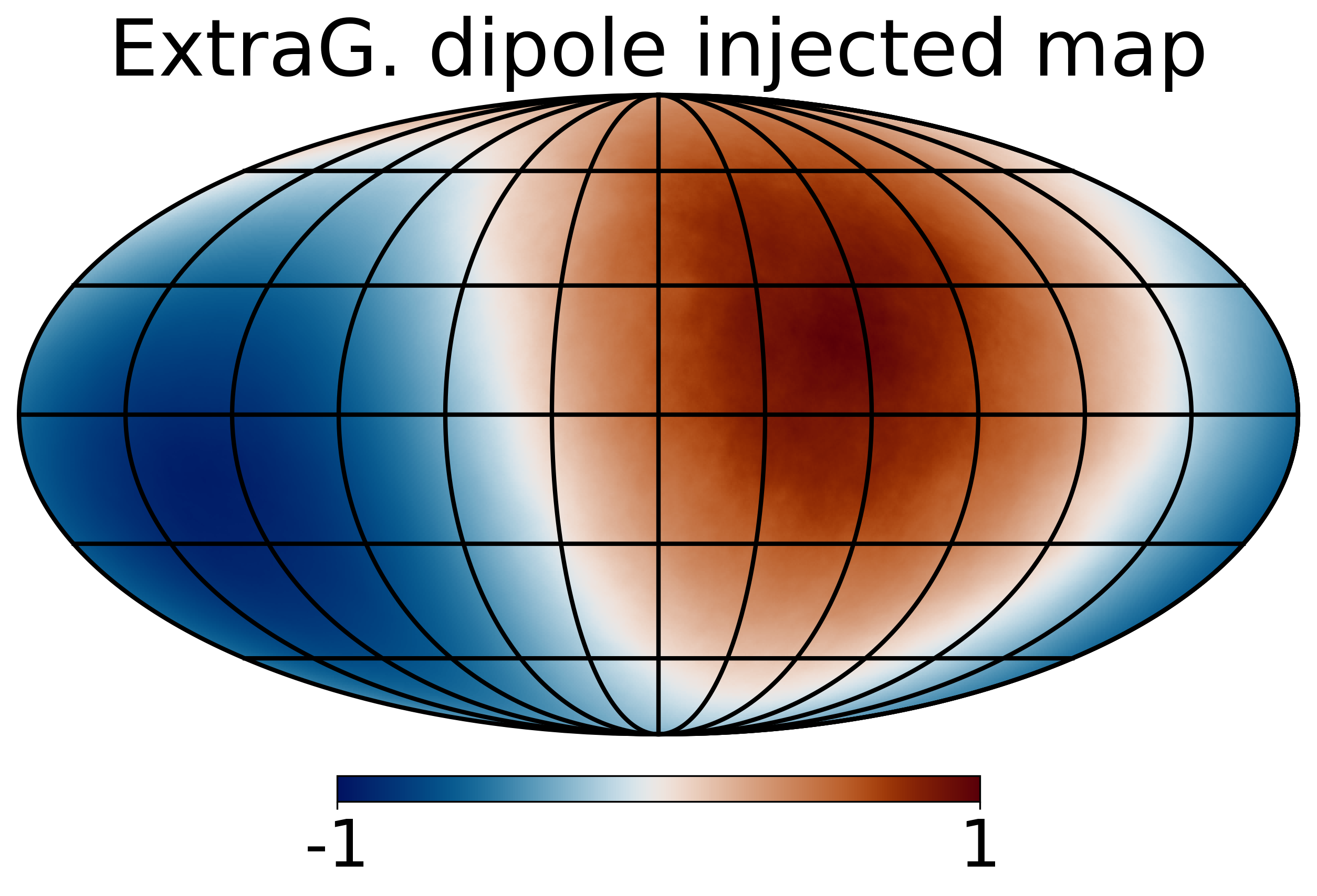}
            \label{fig:Injec_Skymaps_a}
        \end{subfigure}
        
        \vspace{1em}
        \begin{subfigure}[t]{\linewidth}
            \centering
            \includegraphics[width=\linewidth]{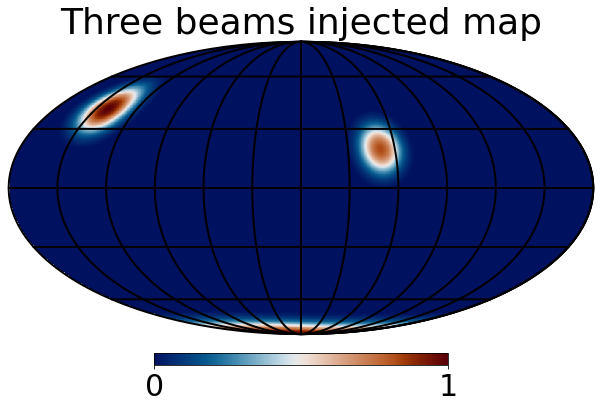}
            \label{fig:Injec_Skymaps_b}
        \end{subfigure}
    \end{minipage}%
    \begin{minipage}[t]{0.45\linewidth}
        \begin{subfigure}[t]{\linewidth}
            \centering
            \includegraphics[width=\linewidth]{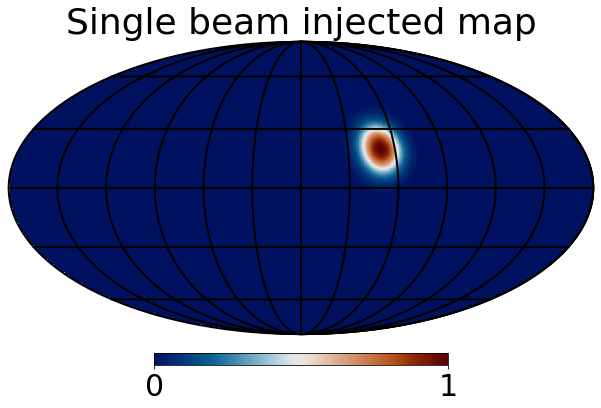}
            \label{fig:Injec_Skymaps_c}
        \end{subfigure}
        
        \vspace{1em}
        \begin{subfigure}[t]{\linewidth}
            \centering
            \includegraphics[width=\linewidth]{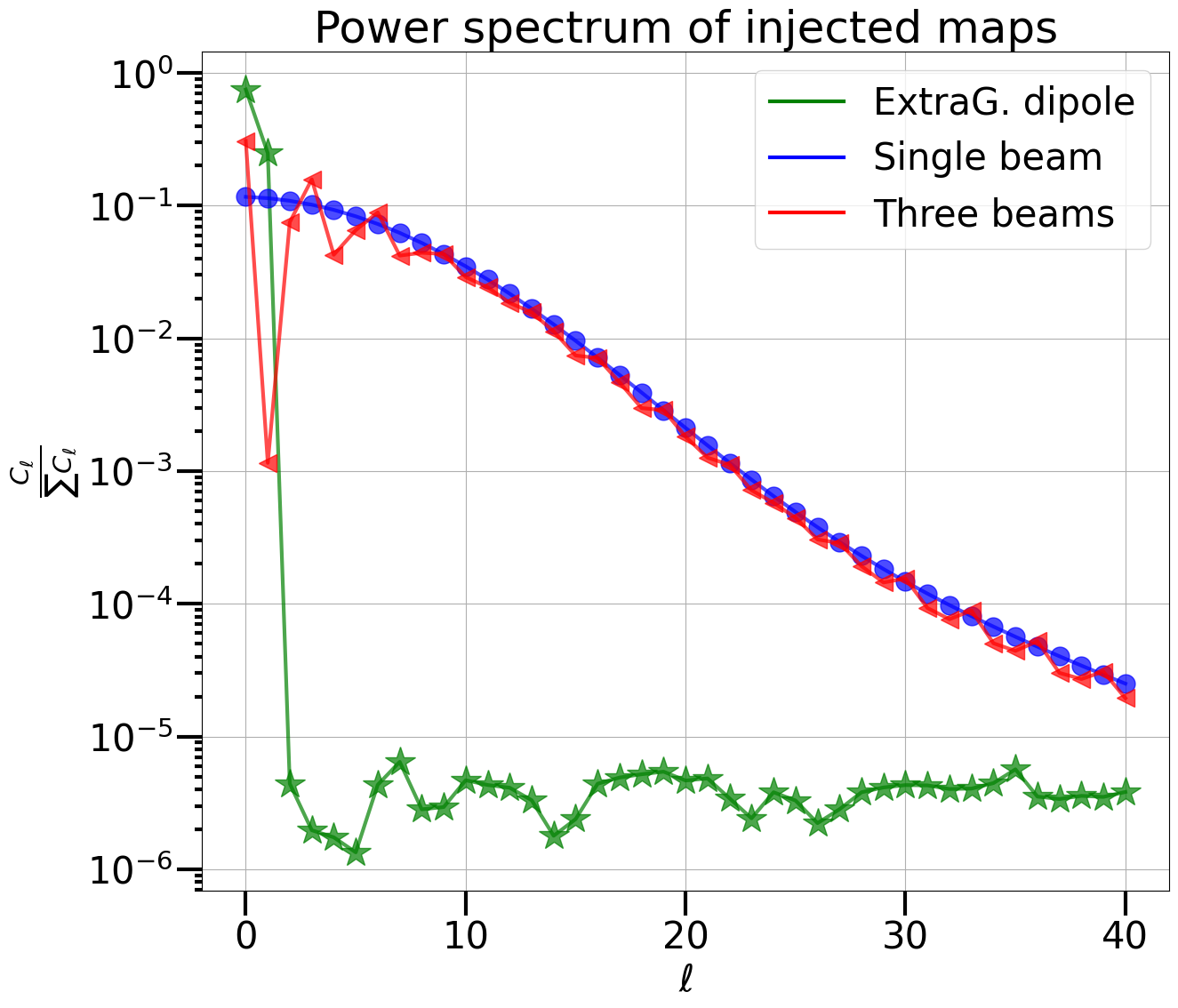}
            \label{fig:Injec_Skymaps_d}
        \end{subfigure}
    \end{minipage}
    \caption{\textbf{Top-left:} panel shows the injected extragalactic dipole skymap. \textbf{Top-right:} shows the single-beam injected skymap. \textbf{Bottom-left:} panel depicts the combined three-beam injected skymap. \textbf{Bottom-right:} panel shows the comparative power spectrum analysis of the three injected skymaps. Note that the maps shown here have a $10^{\circ}$ top-hat smoothing applied.}
    \label{fig:Injec_Skymaps}
\end{figure*}

Forward tracking of UHECRs was performed using CRPropa~3, where 8.5 EeV protons were propagated through the Galaxy's turbulent magnetic field component after being injected inward from an injection sphere. The observer sphere radius $R_{\rm O} $ was set to {1~kpc}, comparable to the size of $L_{\rm {coh}}$. The simulation ends when the particle reaches either the observer sphere or the termination sphere (see Fig.~\ref{fig:cartoons}). The injection sphere was kept at a radius, $R_{\rm I} = $ 30~kpc, and the escape boundary was twice as large as the injection sphere at $R_{\rm T} = $ 60~kpc.

We consider three injection scenarios, namely: 
\vspace{-2mm}
\begin{enumerate}
    \item An {\bf extragalactic dipole} case where we impose an external dipole at $\sqrt{3}$ level at the injection sphere (see \textbf{top-left} skymap Fig.~\ref{fig:Injec_Skymaps}). The value $\sqrt{3}$ comes from the prefactor in  $\delta_{0} = \sqrt{3} \frac{C_{1}}{C_{0}}$. {We adopted a Lambert distribution (isotropic) for every injection point on the injection sphere.}
    \item A {\bf single beam} case where we inject a single $10^\circ$ beam of UHECRs at the position of Centaurus~A ($l = 309.5^{\circ}$, $b = +19.4^{\circ}$) or Cen~A (see \textbf{top-right} skymap Fig.~\ref{fig:Injec_Skymaps}). 
    \item A {\bf three beam} case where we inject three $10^\circ$ beams of UHECRs at the positions of Cen~A, NGC~253 ($l = 97.4^{\circ}$, $b = -87.9^{\circ}$) and M82 ($l = 141.4^{\circ}$, $b = +40.6^{\circ}$)  (see \textbf{bottom-left} skymap Fig.~\ref{fig:Injec_Skymaps}).  {In the three-beam configuration,  two of the beam directions, aligned with the positions of NGC 253 and M82, are assigned identical weights of 35\%, while the third beam direction, aligned with the position of Cen~A, is weighted at 30\%}. Note that the monopole has been removed from the injected skymaps for the single and three-beam cases. {Additionally, we smooth the injected skymaps by $10^{\circ}$.}
\end{enumerate}




The choice of the injected scenarios was motivated by the observations of PAO and TA which showcases hotspots around Cen~A and NGC~253 suggesting these two objects as potential UHECR sources \citep{TA_2014, Auger_2017, Auger_Starburst2018, Auger_2022,Auger_2024}.  The selection of the three injected source scenarios was also motivated by theoretical models \citep{Bell_2022,Taylor_2023} that propose objects within the Council of Giants region as potential sources of UHECRs in our cosmic neighbourhood.

CRPropa~3 and Healpix \citep{Healpix_2005,Healpy_2019} are used to calculate the injected skymaps (see Fig.~\ref{fig:Injec_Skymaps})  with the resolution NSIDE = 512. To calculate the power spectrum of the injected skymaps we used the Anafast function from Healpix\footnote{\href{https://healpix.jpl.nasa.gov/html/intronode6.htm}{Healpix documentation}}. The spherical harmonic coefficient ($\hat{a}_{\ell m}$) is given by:




\begin{align}
\label{eq:alm_def}
\hat{a}_{\ell m} =  \int \rm{d}\Omega \left[\sqrt{\frac{(2\ell + 1)}{4\pi} \frac{(\ell-\textit{m})!}{(\ell + \textit{m})!}} P_{\ell \textit{m}}(\cos\theta)e^{i{\textit m}\phi}\right] \frac{1}{n}\frac{\text{d}n}{\text{d}\Omega} ,
\end{align}

\noindent
here, $\ell = 0....\infty$ is the multipole moment corresponding to angular scales of $180^\circ/\ell$, and $m$ is the azimuthal number ranging from $-\ell$ to $+\ell$. $P_{\ell {m}}(\cos\theta)$ are the associated Legendre polynomials, $\theta$ and $\phi$ are the co-latitude and azimuthal angle, and $(1/{n})(\text{d}n/\text{d}\Omega)$ represents the normalized differential cosmic ray number density per solid angle~\citep{Lang_2021}.

The angular power-spectrum, $C_{\ell}$, at each multipole, $\ell$, can be written in terms of the root-mean-square amplitude of the spherical harmonic coefficients by averaging over the azimuthal number ${m}$ as:  


\begin{align}
    \label{eq:amp}
    C_{\ell}  &= \frac{4\pi}{(2\ell+1)}~{a_\mathrm{rms}^{\ell}}^2  = \frac{4\pi}{(2\ell+1)} \sum_{m=-l}^{l}~| {\hat{a}_\mathrm{\ell \textit{m}}} |^2 .
\end{align}

\noindent
Eq.~\ref{eq:amp} is used to calculate the normalised power spectra for the different injection cases considered (for the case of $L_{\rm{scat}} \approx 80$~kpc), shown in the \textbf{bottom-right} panel in Fig.~\ref{fig:Injec_Skymaps}. In this figure, we show the power spectrum for the {\bf extragalactic dipole} scenario in green, the {\bf single beam} case shown in blue, and the power spectrum of the {\bf three beams} scenario is shown in red. {The power spectra for the injected dipole case shows that the power resides in the monopole and the dipole components. The higher order multipoles are at the noise level.} The power spectra for the injected single beam case looks smooth as the power is derived from a single spot and the rest of the map is flat. In the case of three injected beams the power spectra are more complicated with more structures on large angular scales.

\section{Analysis and results}

\begin{figure*}
    \centering
    \begin{minipage}[t]{0.45\linewidth}
        \begin{subfigure}[t]{\linewidth}
            \includegraphics[width=\linewidth]{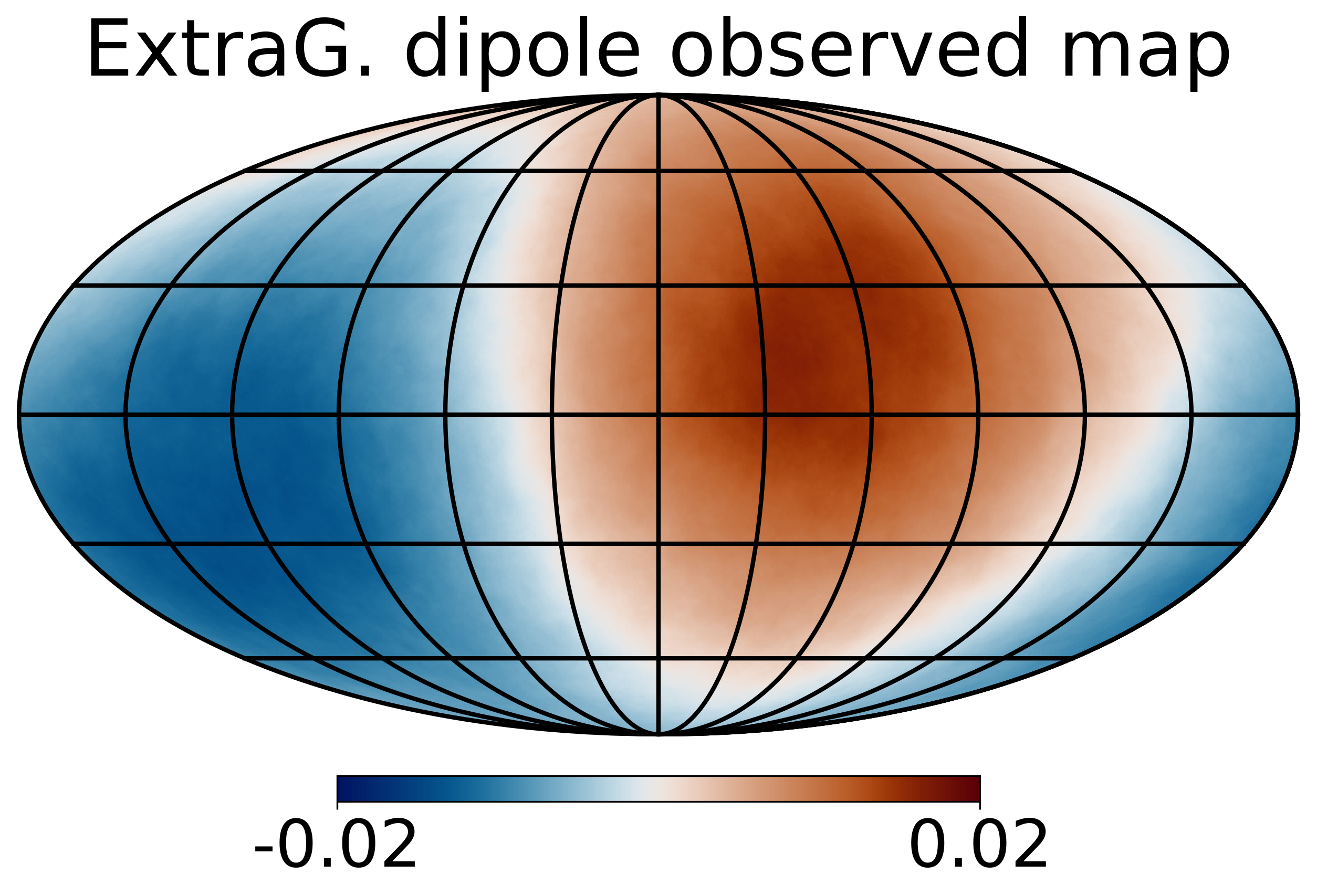}
            \label{fig:Obs_Skymaps_a}
        \end{subfigure}
        
        \vspace{1em}
        \begin{subfigure}[t]{\linewidth}
            \includegraphics[width=\linewidth]{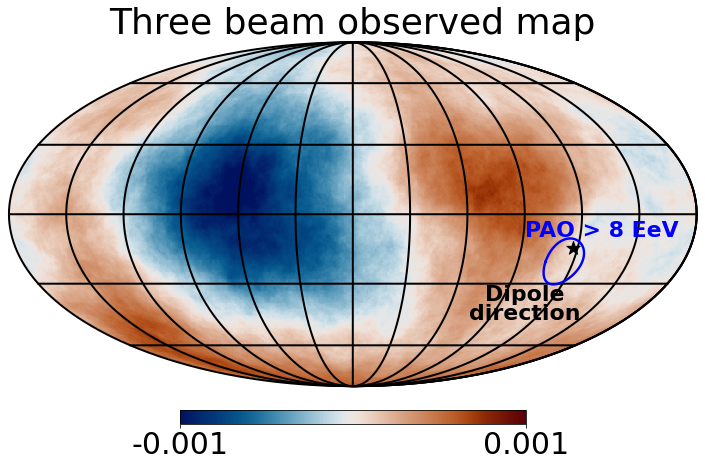}

            \label{fig:Obs_Skymaps_b}
        \end{subfigure}
    \end{minipage}%
    \begin{minipage}[t]{0.45\linewidth}
        \begin{subfigure}[t]{\linewidth}
            \includegraphics[width=\linewidth]{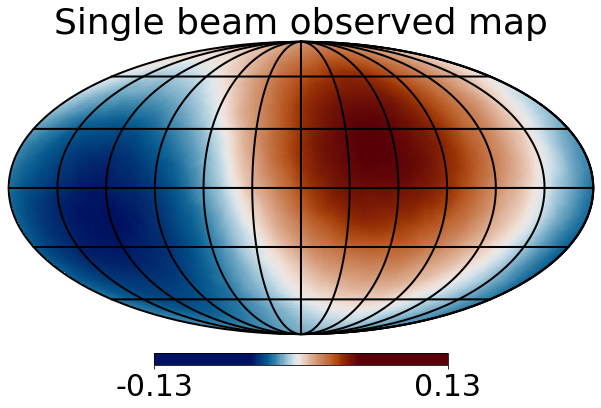}

            \label{fig:Obs_Skymaps_c}
        \end{subfigure}
        
        \vspace{1em}
        \begin{subfigure}[t]{\linewidth}
            \includegraphics[width=\linewidth]{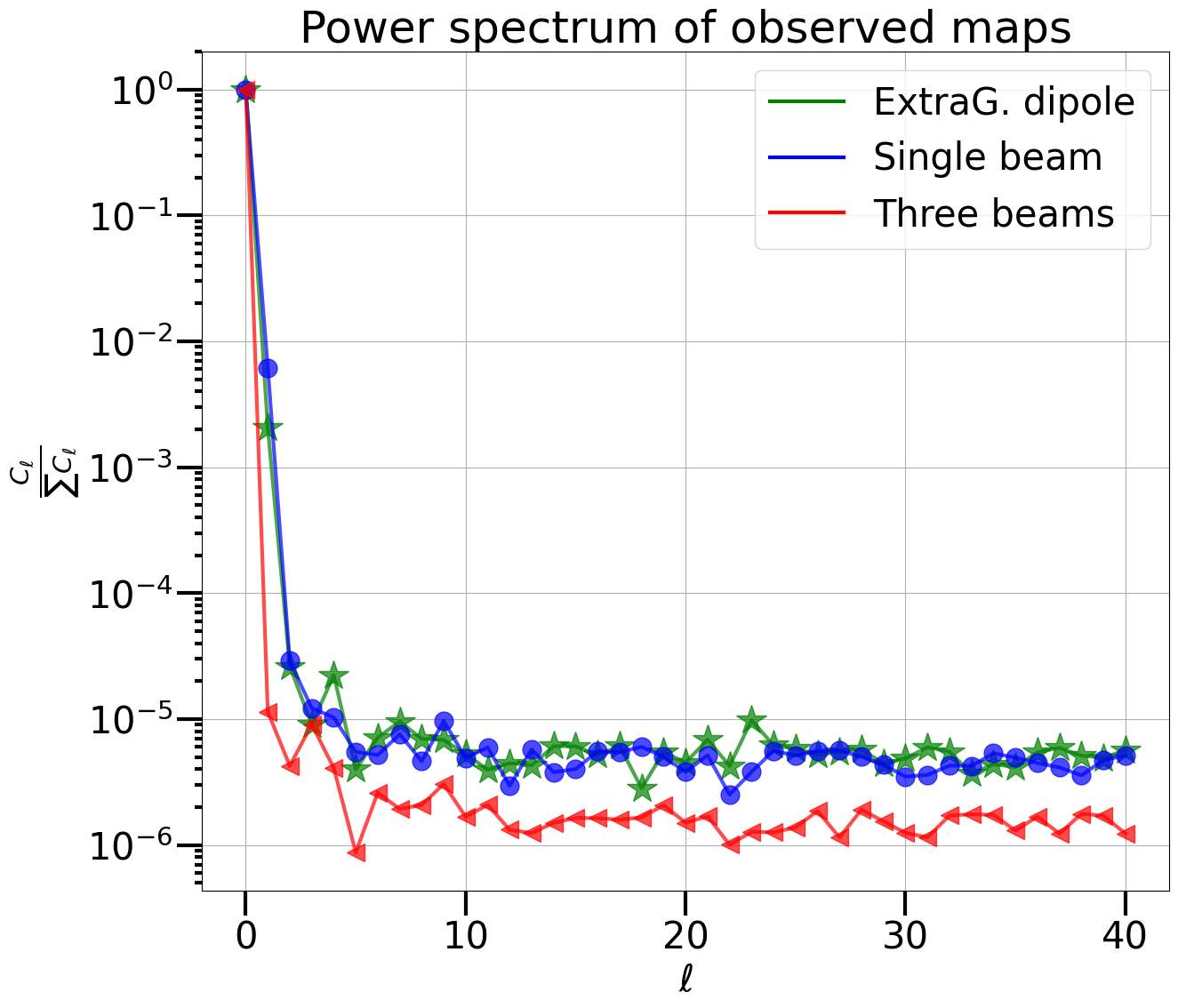}

            \label{fig:Obs_Skymaps_d}
        \end{subfigure}
    \end{minipage}
    \caption{\textbf{Top-left:} observed dipole skymap when an extragalactic dipole is injected with the monopole being removed from the skymaps. \textbf{Top-right:} dipole skymap observed when a single beam is the injected source of UHECRs. \textbf{Bottom-left:} observed dipole skymap for the injected three-beam scenario. The blue star shows the dipole direction from our simulation and the black circle shows the region provided by PAO for rigidities between 8 to 16~EeV~\citep{Auger_2024}. {These skymaps are all produced for $L_{\rm{scat}} \approx 80$~kpc ($B_{\rm{tur}}\approx 7.8~\mu$G) and have a $45^{\circ}$ top-hat smoothing applied, similar to that adopted by the PAO  \citep{Auger_2022,Auger_2024}}. \textbf{Bottom-right:} power spectrum of the observed skymaps for the 3 cases shown above.}    
    \label{fig:Obs_Skymaps}
\end{figure*}

We use the framework mentioned above to calculate the observed sky (see Fig.~\ref{fig:Obs_Skymaps}) for the same NSIDE. In calculating the dipole ($a_\mathrm{rms}^{\ell = 1}$) and quadrupole ($a_\mathrm{rms}^{\ell = 2}$) amplitudes we use two different types of normalisation given by: 
\vspace{-2mm}
\begin{enumerate}
    \item \textbf{Normalisation by monopole:} given by Eq.~\ref{eq:amp_norm_mono}, here we normalise the dipole $a_\mathrm{rms}^{\ell = 1}$ and the quadrupole $a_\mathrm{rms}^{\ell = 2}$ by the monopole amplitude ($a_\mathrm{rms}^{\ell = 0}$).
    \item \textbf{Normalisation by sum of amplitudes:} given by Eq.~\ref{eq:amp_norm_a_all}, here we normalise both the dipole $a_\mathrm{rms}^{\ell = 1}$ and the quadrupole $a_\mathrm{rms}^{\ell = 2}$ by the sum of amplitudes ($\sum\limits_{l=0}\limits^{n} a_\mathrm{rms}^{\ell}$) where $n = 7$. Beyond $\ell = 7$ value, the $C_{\ell}$ values were in the noise domain. 
\end{enumerate}

\noindent
\begin{align}
    \label{eq:amp_norm_mono}
    \delta_{0} = \frac{a_\mathrm{rms}^{l=1}}{a_\mathrm{rms}^{l=0}},  &~
    \epsilon_{0} = \frac{a_\mathrm{rms}^{l=2}}{a_\mathrm{rms}^{l=0}}
\end{align}
\noindent 
\begin{align}
    \label{eq:amp_norm_a_all}
    \delta_{\rm {all}}= \frac{a_\mathrm{rms}^{l=1}}{\sum\limits_{\mathrm {l=0}}\limits^{\mathrm n} a_\mathrm{rms}^{l}}, &~
    \epsilon_{\rm{all}} = \frac{a_\mathrm{rms}^{l=2}}{\sum\limits_{\mathrm {l=0}}\limits^{\mathrm n}a_\mathrm{rms}^{l}}
\end{align}

To estimate the uncertainties we use the same simulation setup as described in Sec.~\ref{sec:sim_setup}, {but for an isotropic distribution} of particles (of the same number as Sec.~\ref{sec:sim_setup} ) through the Galaxy. The injected and observed skymaps obtained are used to determine the dipole and quadrupole amplitudes given by Eq.~\ref{eq:amp_norm_mono} {for this setup, which are random in origin}. {We repeated this $10^4$ times and fit the distribution of the dipole and the quadrupole with a Gaussian. The 99.5\% upper limit on the noise dipole and quadrupole amplitudes from this distribution provide a limit below which any amplitude calculated is considered to be at the noise level.} {Analytically, this estimate on the noise level for each multipole can be estimated by $3\sqrt{2^l/N_{\rm{obs}}}$, where $l$ is the multipole number and $N_{\rm obs}$ is the number of particles in the UHECR skymap.}




The observed skymaps for cases \textit{1,~2,} and \textit{3} are shown in the \textbf{top-right, top-left} and \textbf{bottom-left} panel of Fig.~\ref{fig:Obs_Skymaps}, respectively. Similar to the injected skymaps the NSIDE = 512 and they have been smoothed by a top-hat function with an angle of $45^{\circ}$~\citep{Auger_2022,Auger_2024}. For the case when the diffusive sphere in which the UHECRs propagate is comparable or larger than the scattering length of the UHECRs, a large-scale dipolar skymap is obtained. Similar to the injected case we remove the monopole for the {extragalactic dipole}, single and three-beam observed skymaps. In the three beam case (see \textbf{bottom-left} panel in Fig.~\ref{fig:Obs_Skymaps}) the choice of our weights given to the three sources results in observed dipole direction not very far from what is observed by PAO~\citep{Auger_2024}. The power spectrum for the three cases is shown in the \textbf{bottom-right} panel of Fig.~\ref{fig:Obs_Skymaps}. 


In Fig.~\ref{fig:dipole_quad_supp} we examine the observed dipole and quadrupole amplitudes for the three distinct source scenarios and show the following:
\vspace{-2mm}
\begin{enumerate}
    \item \textbf{Left panel of Fig.~\ref{fig:dipole_quad_supp}:} The observed dipole amplitude, $\delta_{0}$, is given by the hollow markers in red, green, and blue for the extragalactic dipole, single and three-beam cases. The dipole is seen to become gradually suppressed as the scattering length, $L_{\rm{scat}}$, is decreased. 
    The solid markers denote the observed dipole amplitude, $\delta_{\rm{all}}$, colour-coded as above, in this case. {For the single and three beam cases, starting at small $L_{\rm{scat}}$, as $L_{\rm{scat}}$ is increased the dipole amplitude initially rises and then falls. The growth in the dipole amplitude results from a reduction in diffusive "optical depth" of the halo ($R_{\rm I} / L_{\rm{scat}}$). The subsequent decrease in the dipole ($\delta_{\rm{all}}$) at larger $L_{\rm scat}$ values results from power migrating from smaller to larger $\ell$ as $L_{\rm{scat}}$ is increased, and the corresponding probability that particles in the volume scatter before escaping decreases.} In appendix~\ref{appendix_B}, we plot the power spectrum for three different scattering length values in Fig.~\ref{fig:PS_3beam_diff_Lscat} to demonstrate this point. In the extragalactic dipole case, the solid red markers ($\delta_{\rm{all}}$) behave similarly to hollow red markers ($\delta_{0}$), as power at large values of $\ell$ ($>$ 1) is at the noise level. Knowledge of the dipole amplitude alone is insufficient to distinguish between the different source scenarios considered.

    \item \textbf{Right panel of Fig.~\ref{fig:dipole_quad_supp}}. The observed quadrupole amplitude, $\epsilon_{0}$, is given by the hollow markers and  $\epsilon_{\rm{all}}$ is given by the solid markers in red, green, and blue for the extragalactic dipole, single and three-beam cases. {For the extragalactic dipole scenario only a dipole is injected, so no significant quadrupole amplitude is expected to be observed}. A non-zero quadrupole amplitude is only obtained for the single and three beam cases. {For these beam scenarios, starting from small $L_{\rm{scat}}$, as $L_{\rm{scat}}$ is increased the quadrupole amplitude ($\epsilon_{\rm{all}}$) initially rises and then falls. This behaviour occurs for similar reasons as for the dipole behaviour.}
\end{enumerate}

\section{Discussion}
In our work, we have investigated different source scenarios, namely extragalactic dipole and extragalactic beam scenarios. We have investigated how the subsequent observed skymap, for a Galactic plane observer, can be used to distinguish between these different scenarios. One key limitation of our study is that it focused solely on a single realisation of the isotropic turbulent magnetized halo. This assumption was made as the main aim of this work was to study the suppression of the UHECR dipole when propagated through a diffusive sphere with varying diffusion coefficients. 

A further limitation of our work is that we only examine three specific injection cases. Furthermore, for the beam scenarios considered, we have restricted the beam to an ad-hoc size of $10^{\circ}$, and we did not scan over different weights for the different beams. Additionally, we restricted the position of these beams to the potential UHECR source candidates that lie on the Council of Giants. For a realistic comparison with UHECR data and exploration of all possibilities, a scan of different source positions, beam sizes, relative beam weights, and combinations with isotropic background should also be examined. However, this is outside the scope of this work.

\begin{figure*}
    \centering
    \begin{minipage}[c]{0.49\linewidth}
        \centering
        \includegraphics[width=\linewidth]{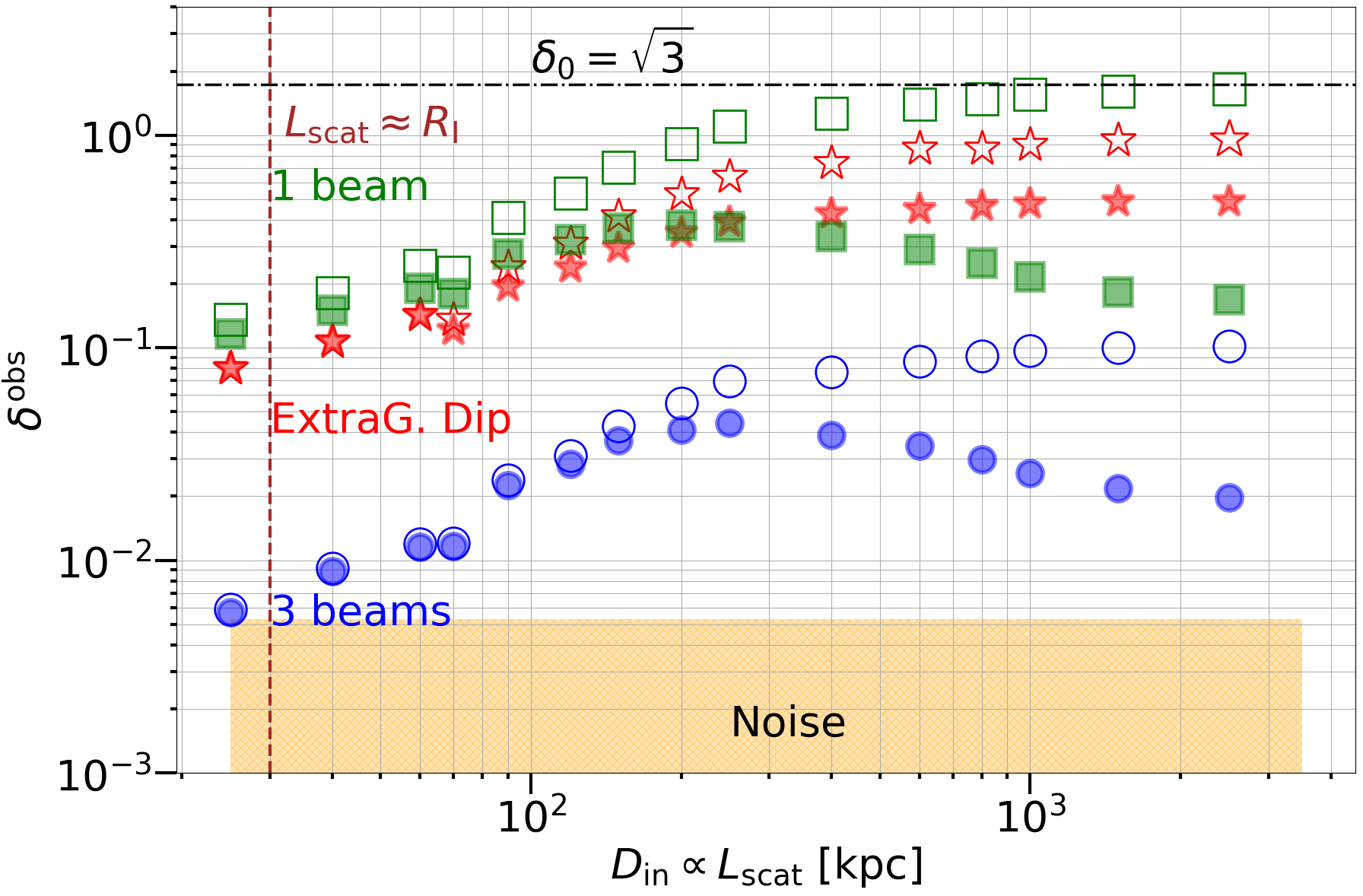}

         \label{dipole_quad_supp_a}

    \end{minipage}%
    \begin{minipage}[c]{0.49\linewidth}
        \centering
            \includegraphics[width=\linewidth]{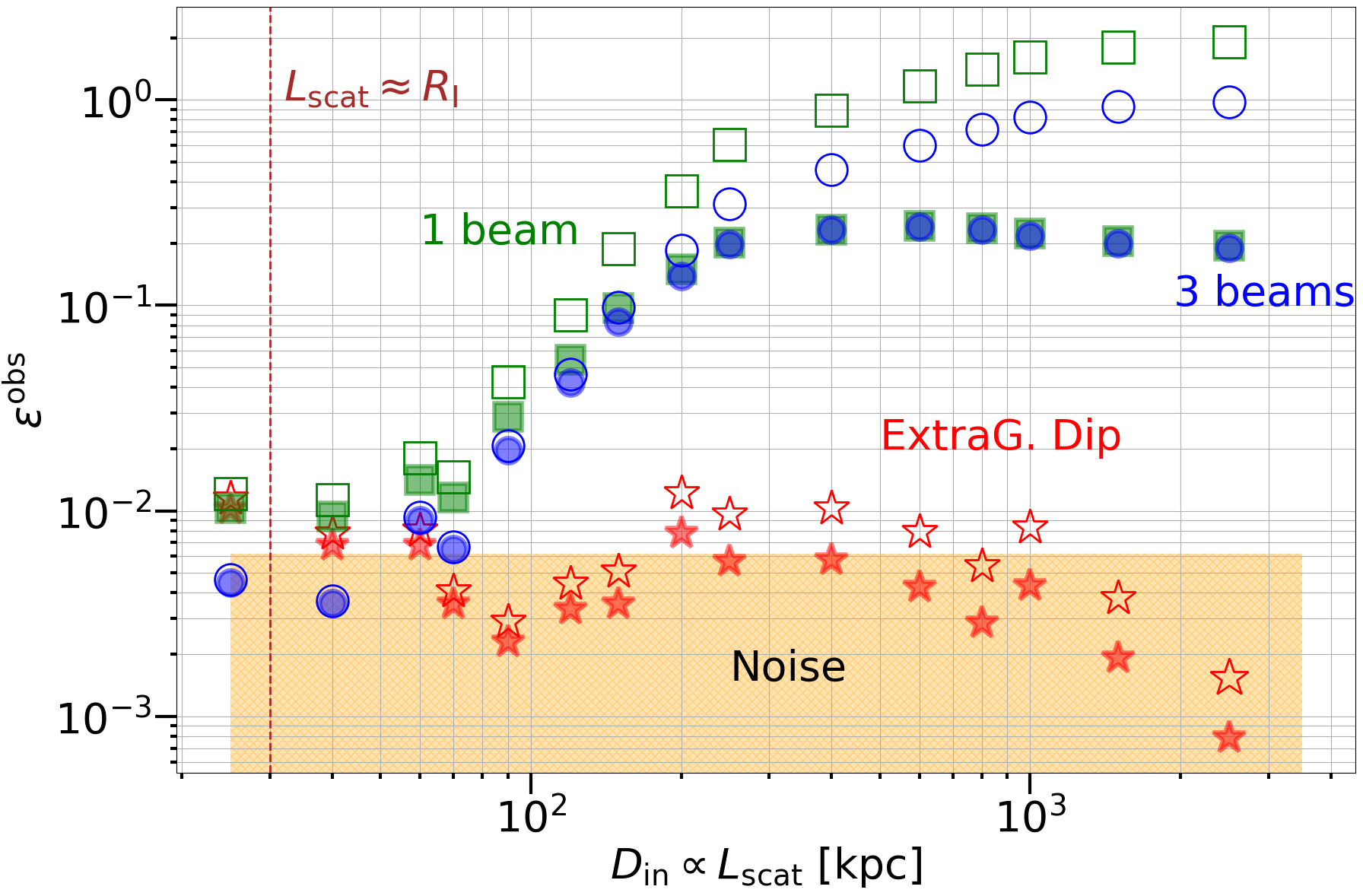}

        \label{dipole_quad_supp_b}
    \end{minipage}
    \caption{\textbf{Left-hand panel:} dipole amplitude $\delta^{\rm{obs}}$ as a function of scattering length for the three cases: \textit{1)} extragalactic dipole shown in red ($\star$), \textit{2)} single-beam case shown in green ($\blacksquare$) and \textit{3)} three-beam case shown in blue (\tikzcircle[black,fill = black]{2pt}). The solid and hollow markers indicate $\delta_{\rm{all}}$ and $\delta_{0}$, respectively, for all three cases. \textbf{Right-hand panel:} quadrupole amplitude for the three cases as on the left figure with the same colour code, here $\epsilon_{\rm{all}}$ and $\epsilon_{0}$ are shown by solid and hollow markers again. The orange band represents the {$99.5\%$ confidence region for the noise}. {The vertical dashed line indicates where $L_{\rm{scat}} \approx R_{\rm I}$.}}
    \label{fig:dipole_quad_supp}
\end{figure*}

The specific three-beam case we investigate gives a relatively weak dipole amplitude and a relatively strong quadrupole amplitude (see Fig.~\ref{fig:dipole_quad_supp}). This combination is challenging to reconcile with PAO data, which shows $\delta^{\rm{obs}} = \rm {0.074}^{+0.010}_{-0.008}$ and $\epsilon_{obs} < 0.05$ for $E \geq 8$~EeV~\citep{Auger_2024}. For all $D_\mathrm{in}$ values that give a strong enough dipole amplitude, the quadrupole amplitude is significantly larger than the upper limit obtained by PAO. It is possible to reduce the quadrupole amplitude by adding an isotropic background of UHECRs to this three-beam case, but this also decreases the dipole amplitude. This shows the strength of combining the dipole and quadrupole amplitudes for investigations in specific source scenarios. However, given that we only consider one specific three-beam scenario, no strong conclusion can be drawn from comparing this investigation with PAO data. 

\textcolor{red}{
} 

Our work is complementary to previous papers where the authors neglected the effect of turbulent magnetic fields inside the Galaxy, with the focus instead being on the effect of EGMF on the UHECR dipole \citep{Harari_2014, Lang_2021}. We find the dipole amplitude to decrease with increasing turbulent magnetic field strength which seems to be consistent with recent work by \citet{Bister_2024}. Complementary to our work, \citet{Bakalov_2023} carried out a phenomenological study of the change of the UHECR dipole amplitude and direction in the presence of specific GMF models. They investigated the dependence on the mass of UHECRs, assuming an extragalactic dipole distribution.

A natural succession to this work would be to examine the dipole and quadrupole behaviour in different realisations of turbulent magnetic fields, and different compositions of UHECR species \citep{Auger_Prime_2016,Auger_Prime_2019}, and the addition of large-scale structured Galactic magnetic fields along with extragalactic magnetic fields.

\section{Conclusions}

Recent observations of emission from the halo of nearby galaxies suggest that a considerable thermal gas pressure fills galactic haloes out to distances beyond 100~kpc from the galactic centres~\citep{2022MNRAS.511..843M,2022ApJ...928...14B,2024A&A...690A.267Z}. Here, we consider the consequences of these observations on extragalactic UHECR propagation through the Milky Way's {giant} Galactic halo, assuming that the halo also possesses a considerable magnetisation fraction. We focus on the effects of a turbulent magnetic field in the halo. To explore a range of possibilities in this direction, we consider three distinct injection scenarios, namely: $\textit {1})$ an extragalactic dipole, $\textit{2}$) a single beam (a small spread of arrival directions around a single direction) and $\textit{3}$) three beams. 

We find that a dipolar skymap is found, regardless of which of the actual three injection scenarios we consider, provided that the particle scattering length is comparable to the halo size. Therefore, the observation of a UHECR dipole, as reported by the Pierre Auger Collaboration~\citep{Auger_2017, Auger_2018, Auger_2024}, is by itself insufficient to reveal the nature of its origin.

Furthermore, for all three scenarios considered, the dipole amplitude was less suppressed when the scattering length in the halo (which scales with rigidity) was increased. The dipole amplitude should, therefore, increase with UHECR rigidity (consistent with recent PAO results~\citep{Auger_2018, Auger_2024}), irrespective of the extragalactic UHECR angular distribution's energy dependence. Likewise, the quadrupole amplitude increases rapidly with scattering length for the two extragalactic beam cases considered.

Beyond this scattering length dependence, we show that the strength of the quadrupolar component can be used to distinguish between our different dipole origin scenarios. If the observed dipole arises due to an extragalactic dipole, a significant quadrupole moment is not observed. However, a significant quadrupole moment is observed for the single-beam and three-beam cases. 




\section*{Data availability}
We make use of the HEALPix~\citep{Healpix_2005}, Healpy~\citep{Healpy_2019}, Matplotlib~\citep{Matplotlib} and Numpy~\citep{Numpy} packages. The code used for the cosmic-ray propagation was CRPropa~3~\citep{CRPropa_2022}\footnote{https://crpropa.github.io/CRPropa3/index.html}, which is also a publicly available software. The codes used for this work can be made available to the corresponding author upon request. 

\section*{Acknowledgements}
VS and AT acknowledge support from DESY (Zeuthen, Germany), a member of the Helmholtz Association HGF. AvV acknowledges support from Khalifa University’s internal FSU-2022-025 and RIG-2024-047 grants. VS would like to thank Clive Dickinson and the STFC Consolidated Grant (ST/P000649/1) for supporting part of the work carried out at the University of Manchester. The authors would also like to thank James Matthews and Damiano Fiorillo for useful discussions. 


%

\bibliographystyle{mnras}
\bibliography{references.bib}

\begin{thebibliography}{}
\makeatletter
\relax
\def\mn@urlcharsother{\let\do\@makeother \do\$\do\&\do\#\do\^\do\_\do\%\do\~}
\def\mn@doi{\begingroup\mn@urlcharsother \@ifnextchar [ {\mn@doi@} {\mn@doi@[]}}
\def\mn@doi@[#1]#2{\def\@tempa{#1}\ifx\@tempa\@empty \href {http://dx.doi.org/#2} {doi:#2}\else \href {http://dx.doi.org/#2} {#1}\fi \endgroup}
\def\mn@eprint#1#2{\mn@eprint@#1:#2::\@nil}
\def\mn@eprint@arXiv#1{\href {http://arxiv.org/abs/#1} {{\tt arXiv:#1}}}
\def\mn@eprint@dblp#1{\href {http://dblp.uni-trier.de/rec/bibtex/#1.xml} {dblp:#1}}
\def\mn@eprint@#1:#2:#3:#4\@nil{\def\@tempa {#1}\def\@tempb {#2}\def\@tempc {#3}\ifx \@tempc \@empty \let \@tempc \@tempb \let \@tempb \@tempa \fi \ifx \@tempb \@empty \def\@tempb {arXiv}\fi \@ifundefined {mn@eprint@\@tempb}{\@tempb:\@tempc}{\expandafter \expandafter \csname mn@eprint@\@tempb\endcsname \expandafter{\@tempc}}}

\bibitem[\protect\citeauthoryear{Aab et~al.,}{Aab et~al.}{2016}]{Auger_Prime_2016}
Aab A.,  et~al., 2016 (\mn@eprint {arXiv} {1604.03637})

\bibitem[\protect\citeauthoryear{Aab et~al.}{Aab et~al.}{2017}]{Auger_2017}
Aab A.,  et~al., 2017, \mn@doi [Science] {10.1126/science.aan4338}, 357, 1266

\bibitem[\protect\citeauthoryear{Aab et~al.,}{Aab et~al.}{2018a}]{Auger_Starburst2018}
Aab A.,  et~al., 2018a, \mn@doi [\apj] {10.3847/2041-8213/aaa66d}, 853, L29

\bibitem[\protect\citeauthoryear{Aab et~al.}{Aab et~al.}{2018b}]{Auger_2018}
Aab A.,  et~al., 2018b, \mn@doi [ApJ] {10.3847/1538-4357/aae689}, 868, 4

\bibitem[\protect\citeauthoryear{Abbasi et~al.,}{Abbasi et~al.}{2014}]{TA_2014}
Abbasi R.~U.,  et~al., 2014, \mn@doi [\apj] {10.1088/2041-8205/790/2/l21}, 790, L21

\bibitem[\protect\citeauthoryear{Abbasi et~al.}{Abbasi et~al.}{2018}]{TelescopeArray:2018rtg}
Abbasi R.~U.,  et~al., 2018, \mn@doi [ApJ] {10.3847/1538-4357/aac9c8}, 862, 91

\bibitem[\protect\citeauthoryear{Abbasi et~al.}{Abbasi et~al.}{2020}]{Abbasi_2020}
Abbasi R.~U.,  et~al., 2020, \mn@doi [ApJ Lett.] {10.3847/2041-8213/aba0bc}, 898, L28

\bibitem[\protect\citeauthoryear{Abdul~Halim et~al.}{Abdul~Halim et~al.}{2023}]{PierreAuger:2023mvf}
Abdul~Halim A.,  et~al., 2023, \mn@doi [PoS] {10.22323/1.444.0521}, ICRC2023, 521

\bibitem[\protect\citeauthoryear{Abdul~Halim et~al.}{Abdul~Halim et~al.}{2024}]{Auger_2024}
Abdul~Halim A.,  et~al., 2024, \mn@doi [ApJ] {10.3847/1538-4357/ad843b}, 976, 48

\bibitem[\protect\citeauthoryear{Abreu et~al.}{Abreu et~al.}{2022}]{Auger_2022}
Abreu P.,  et~al., 2022, \mn@doi [ApJ] {10.3847/1538-4357/ac7d4e}, 935, 170

\bibitem[\protect\citeauthoryear{Allard, Aublin, Baret  \& Parizot}{Allard et~al.}{2022}]{Allard:2021ioh}
Allard D.,  Aublin J.,  Baret B.,   Parizot E.,  2022, \mn@doi [A\&A] {10.1051/0004-6361/202142491}, 664, A120

\bibitem[\protect\citeauthoryear{Alves~Batista, Dundovic, Erdmann, Kampert, Kuempel  et~al.}{Alves~Batista et~al.}{2016}]{CRPropa_2016}
Alves~Batista A.,  Dundovic A.,  Erdmann M.,  Kampert K.-H.,  Kuempel D.,   et~al., 2016, \mn@doi [\jcap] {10.1088/1475-7516/2016/05/038}, 2016, 038–038

\bibitem[\protect\citeauthoryear{Alves~Batista, Tjus, Dörner, Dundovic, Eichmann  et~al.}{Alves~Batista et~al.}{2022}]{CRPropa_2022}
Alves~Batista R.,  Tjus J.~B.,  Dörner J.,  Dundovic A.,  Eichmann B.,   et~al., 2022, \mn@doi [\jcap] {10.1088/1475-7516/2022/09/035}, 2022, 035

\bibitem[\protect\citeauthoryear{Bakalová, Vícha  \& Trávníček}{Bakalová et~al.}{2023}]{Bakalov_2023}
Bakalová A.,  Vícha J.,   Trávníček P.,  2023, \mn@doi [\jcap] {10.1088/1475-7516/2023/12/016}, 2023, 016

\bibitem[\protect\citeauthoryear{Bell \& Matthews}{Bell \& Matthews}{2022}]{Bell_2022}
Bell A.~R.,  Matthews J.~H.,  2022, \mn@doi [\mnras] {10.1093/mnras/stac031}, 511, 448

\bibitem[\protect\citeauthoryear{Bister \& Farrar}{Bister \& Farrar}{2024}]{Bister:2023icg}
Bister T.,  Farrar G.~R.,  2024, \mn@doi [ApJ] {10.3847/1538-4357/ad2f3f}, 966, 71

\bibitem[\protect\citeauthoryear{Bister, Farrar  \& Unger}{Bister et~al.}{2024}]{Bister_2024}
Bister T.,  Farrar G.~R.,   Unger M.,  2024, \mn@doi [ApJL] {10.3847/2041-8213/ad856f}, 975, L21

\bibitem[\protect\citeauthoryear{{Bregman}, {Hodges-Kluck}, {Qu}, {Pratt}, {Li}  \& {Yun}}{{Bregman} et~al.}{2022}]{2022ApJ...928...14B}
{Bregman} J.~N.,  {Hodges-Kluck} E.,  {Qu} Z.,  {Pratt} C.,  {Li} J.-T.,   {Yun} Y.,  2022, \mn@doi [\apj] {10.3847/1538-4357/ac51de}, \href {https://ui.adsabs.harvard.edu/abs/2022ApJ...928...14B} {928, 14}

\bibitem[\protect\citeauthoryear{Castellina et~al.,}{Castellina et~al.}{2019}]{Auger_Prime_2019}
Castellina A.,  et~al., 2019, \mn@doi [EPJ Web Conf.] {10.1051/epjconf/201921006002}, 210, 06002

\bibitem[\protect\citeauthoryear{Ding, Globus  \& Farrar}{Ding et~al.}{2021}]{Ding:2021emg}
Ding C.,  Globus N.,   Farrar G.~R.,  2021, \mn@doi [ApJ Lett.] {10.3847/2041-8213/abf11e}, 913, L13

\bibitem[\protect\citeauthoryear{Eichmann, Kachelrie\ss{}  \& Oikonomou}{Eichmann et~al.}{2022}]{Eichmann:2022ias}
Eichmann B.,  Kachelrie\ss{} M.,   Oikonomou F.,  2022, \mn@doi [JCAP] {10.1088/1475-7516/2022/07/006}, 07, 006

\bibitem[\protect\citeauthoryear{{Faerman}, {Sternberg}  \& {McKee}}{{Faerman} et~al.}{2020}]{2020ApJ...893...82F}
{Faerman} Y.,  {Sternberg} A.,   {McKee} C.~F.,  2020, \mn@doi [\apj] {10.3847/1538-4357/ab7ffc}, \href {https://ui.adsabs.harvard.edu/abs/2020ApJ...893...82F} {893, 82}

\bibitem[\protect\citeauthoryear{Gorski, Hivon, Banday, Wandelt, Hansen, Reinecke  \& Bartelmann}{Gorski et~al.}{2005}]{Healpix_2005}
Gorski K.~M.,  Hivon E.,  Banday A.~J.,  Wandelt B.~D.,  Hansen F.~K.,  Reinecke M.,   Bartelmann M.,  2005, \mn@doi [\apj] {10.1086/427976}, 622, 759

\bibitem[\protect\citeauthoryear{Harari, Mollerach  \& Roulet}{Harari et~al.}{2014}]{Harari_2014}
Harari D.,  Mollerach S.,   Roulet E.,  2014, \mn@doi [PRD] {10.1103/physrevd.89.123001}, 89

\bibitem[\protect\citeauthoryear{{Harris} et~al.,}{{Harris} et~al.}{2020}]{Numpy}
{Harris} C.~R.,  et~al., 2020, \mn@doi [\nat] {10.1038/s41586-020-2649-2}, 585, 357

\bibitem[\protect\citeauthoryear{Heesen, O’Sullivan, Brüggen, Basu, Beck, Seta  \& Carretti}{Heesen et~al.}{2023}]{Heesen_2023}
Heesen V.,  O’Sullivan S.~P.,  Brüggen M.,  Basu A.,  Beck R.,  Seta A.,   Carretti E.,  2023, \mn@doi [A\&A] {10.1051/0004-6361/202346008}, 670, L23

\bibitem[\protect\citeauthoryear{Hopkins, Butsky, Panopoulou, Ji, Quataert, Faucher-Giguère  \& Kereš}{Hopkins et~al.}{2022}]{Hopkins_2022}
Hopkins P.~F.,  Butsky I.~S.,  Panopoulou G.~V.,  Ji S.,  Quataert E.,  Faucher-Giguère C.-A.,   Kereš D.,  2022, \mn@doi [Monthly Notices of the Royal Astronomical Society] {10.1093/mnras/stac1791}, 516, 3470–3514

\bibitem[\protect\citeauthoryear{{Hunter}}{{Hunter}}{2007}]{Matplotlib}
{Hunter} J.~D.,  2007, \mn@doi [Comput. in Sci. and Eng.] {10.1109/MCSE.2007.55}, 9, 90

\bibitem[\protect\citeauthoryear{Lang, Taylor  \& de Souza}{Lang et~al.}{2021}]{Lang_2021}
Lang R.~G.,  Taylor A.~M.,   de Souza V.,  2021, \mn@doi [PRD] {10.1103/physrevd.103.063005}, 103

\bibitem[\protect\citeauthoryear{{Martynenko}}{{Martynenko}}{2022}]{2022MNRAS.511..843M}
{Martynenko} N.,  2022, \mn@doi [\mnras] {10.1093/mnras/stac164}, \href {https://ui.adsabs.harvard.edu/abs/2022MNRAS.511..843M} {511, 843}

\bibitem[\protect\citeauthoryear{Mollerach \& Roulet}{Mollerach \& Roulet}{2019}]{Mollerach:2019wne}
Mollerach S.,  Roulet E.,  2019, \mn@doi [Phys. Rev. D] {10.1103/PhysRevD.99.103010}, 99, 103010

\bibitem[\protect\citeauthoryear{Mollerach \& Roulet}{Mollerach \& Roulet}{2024}]{Mollerach:2024sjd}
Mollerach S.,  Roulet E.,  2024, \mn@doi [Phys. Rev. D] {10.1103/PhysRevD.110.063030}, 110, 063030

\bibitem[\protect\citeauthoryear{Taylor, Matthews  \& Bell}{Taylor et~al.}{2023}]{Taylor_2023}
Taylor A.~M.,  Matthews J.~H.,   Bell A.~R.,  2023, \mn@doi [\mnras] {10.1093/mnras/stad1716}, 524, 631–642

\bibitem[\protect\citeauthoryear{{Zhang} et~al.,}{{Zhang} et~al.}{2024}]{2024A&A...690A.267Z}
{Zhang} Y.,  et~al., 2024, \mn@doi [\aap] {10.1051/0004-6361/202449412}, \href {https://ui.adsabs.harvard.edu/abs/2024A&A...690A.267Z} {690, A267}

\bibitem[\protect\citeauthoryear{{Zirakashvili}}{{Zirakashvili}}{2005}]{Zirakashvili_2005}
{Zirakashvili} V.~N.,  2005, \mn@doi [Int. J. of Modern Phys. A] {10.1142/S0217751X05030314}, \href {https://ui.adsabs.harvard.edu/abs/2005IJMPA..20.6858Z} {20, 6858}

\bibitem[\protect\citeauthoryear{{Zonca}, {Singer}, {Lenz}, {Reinecke}, {Rosset}, {Hivon}  \& {Gorski}}{{Zonca} et~al.}{2019}]{Healpy_2019}
{Zonca} A.,  {Singer} L.,  {Lenz} D.,  {Reinecke} M.,  {Rosset} C.,  {Hivon} E.,   {Gorski} K.,  2019, \mn@doi [J. of Open Source Software] {10.21105/joss.01298}, 4, 1298

\makeatother
\end{thebibliography}

\appendix

\section{Power spectrum for different $L_{\rm{scat}}$ values}
\label{appendix_B}

\begin{figure}
    \centering
    \includegraphics[width=1\linewidth]{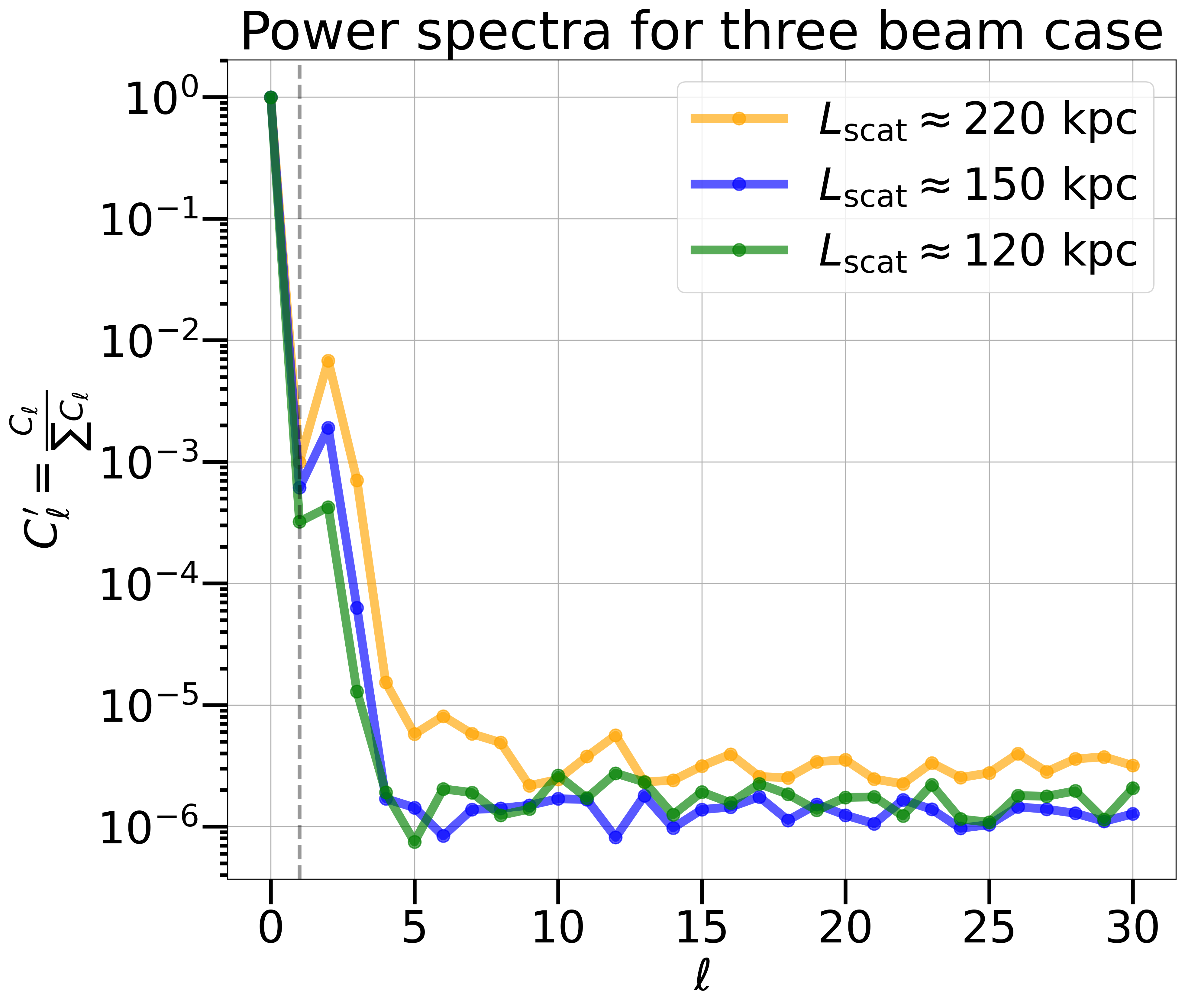}
    \caption{Power spectra for three-beam case with different scattering length values.}
    \label{fig:PS_3beam_diff_Lscat}
\end{figure}

We plot the power spectrum from the observed skymaps for three different values of $L_{\rm {scat}}$ in Fig.~\ref{fig:PS_3beam_diff_Lscat} for the three beam case with $L_{\rm{coh}} \approx$~1~kpc, most of the power sits in the monopole, dipole and quadrupole. 

\end{document}